\documentclass[12pt]{iopart}

\usepackage[english]{babel}
\usepackage{graphicx}
\usepackage{subfigure}
\usepackage{bm}
\usepackage{iopams}
 \expandafter\let\csname equation*\endcsname\relax
  \expandafter\let\csname endequation*\endcsname\relax
\usepackage{amsmath,amssymb,amsfonts}

\usepackage{verbatim}

\begin{document}

\title[Optical evidence of quantum rotor orbital excitations in orthorhombic manganites]{Optical evidence of quantum rotor orbital excitations in orthorhombic manganites}

\author{N N Kovaleva$^{1,2,3}$, K I Kugel$^{2,4}$, Z Pot\r u\v cek$^{5}$, N. S. Goryachev$^{6}$, O E Kusmartseva$^2$, Z Bryknar$^{5}$, V A Trepakov$^{7,3}$, E~I~Demikhov$^1$, A Dejneka$^3$, F V Kusmartsev$^2$ and~ A~ M~ Stoneham$^8$\footnote{Deceased.}}

\address{$^1$ Lebedev Physical Institute, Russian Academy of Sciences, Moscow, Leninsky prosp. 53, 119991, Russia}
\address{$^2$ Department of Physics, Loughborough University, Loughborough, LE11 3TU, UK}
\address{$^3$ Institute of Physics, Academy of Sciences of the Czech Republic, Prague, Na~ Slovance 2, 18221, Czech Republic}
\address{$^4$ Institute for Theoretical and Applied Electrodynamics, Russian Academy of Sciences, Moscow, Izhorskaya street 13/19, 125412, Russia}
\address{$^5$ Czech Technical University, Prague, Trojanova 13, 12000, Czech Republic}
\address{$^6$ Institute of Problems in Chemical Physics, Russian Academy of Sciences, Chernogolovka, Academician Semenov avenue 1, 142432, Russia}
\address{$^7$ A F Ioffe Physical-Technical Institute, Russian Academy of Sciences, St. Petersburg, Politekhnicheskaya street 26, 194021, Russia}
\address{$^8$ London Centre for Nanotechnology, University College London, London, 17-19 Gordon Street, WC1H OAH, UK}

\ead{nkovaleva@sci.lebedev.ru}

\date{\today}

\begin{abstract}
In magnetic compounds with Jahn--–Teller (JT) ions (such as Mn$^{3+}$ or Cu$^{2+}$), the ordering of the electron or hole orbitals is associated with cooperative lattice distortions. There the role of JT effect, although widely recognised, is still elusive in the ground state properties. We suggest that, in these materials, there exist elementary excitations whose energy spectrum is described in terms of the total angular momentum eigenstates and is quantised as in quantum rotors found in JT centers. We observed features originating from these excitations in the optical spectra of a model compound LaMnO$_3$ using ellipsometry technique. They appear clearly as narrow sidebands accompanying the electron transition between the JT split orbitals on neighbouring Mn$^{3+}$ ions, strongly influenced by anisotropic spin correlations. We present these results together with new experimental data on photoluminescence and its kinetics found in LaMnO$_3$, which lend additional support to the ellipsometry implying the existence of the quantum rotor orbital excitations. We note that the discovered elementary excitations of quantum rotors may play an important role in many unusual properties observed in these materials upon doping, such as high-temperature superconductivity and colossal magnetoresistance. 
\end{abstract}



\submitto{\JPCM}

\maketitle


\section{Introduction}

Orbital ordering phenomena in correlated electron systems with electron--phonon interaction remains an actively developing field of research despite many years of scientific effort. In magnetic compounds with Jahn--Teller ions (such as Mn$^{3+}$ or Cu$^{2+}$), the ordering of the $e_g$ electron or hole orbitals leads not only to cooperative lattice distortions, but also determines their magnetic properties. In addition to a more traditional electron--lattice Jahn-Teller (JT) mechanism of orbital ordering \cite{JahnTeller,KaplVekh_book,Bersuker_book}, a purely electronic superexchange (SE) mechanism can also lead to the formation of the orbital and magnetic ordering \cite{Gooden_book,KugelKhomskii,Tokura}. 

Orthorhombic rare-earth (RE) manganites REMnO$_3$ (RE=La,Pr,Nd,Sm,Eu,Gd,Tb,Dy) are regarded to be model compounds in orbital physics: these systems are perovskites with the Mn$^{3+}$ ions having the $t^3_{2g}e^1_g$ electronic configuration. Their high-temperature phase is nearly cubic, and at the temperature of $T_{\rm OO}$, they undergo a transition into the orthorhombic orbitally-ordered phase with an antiferrodistorsive ordering of MnO$_6$ octahedra [see Fig. 1(a,b)]. Recent {\it ab initio} studies demonstrate that the SE interaction alone cannot initiate the transition at $T_{\rm OO}$ \cite{Pavarini}. Another argument supporting the importance of the JT mechanism comes from the fact that  the SE favours the occupation of the $e_g$ orbital with the orbital angle $\theta=90^{\circ}$. In fact, the orbital angle determined from the neutron scattering for LaMnO$_3$ is $\theta \sim 108^{\circ}$ \cite{Rodriguez}, and this difference further increases in the manganites with smaller ionic radius to $\theta \sim114^{\circ}$ for TbMnO$_3$ \cite{Kim}. This indicates that the nature of the $e_g$ electron ground state in orthorhombic manganites is more pertinent to the JT ground state in the ''Mexican hat'' potential profile at each Mn$^{3+}$ site [see Fig. 1(c)] \cite{O'Brien}. Here the ground state may be of infinite degeneracy in the $dynamic$ limit, otherwise the $e_g$ electron is localised in one of the discrete minima created by the lattice anharmonicity and associated with tetragonal $static$ lattice distortion.  

Recently the changes  in the orbital ordering observed in X-ray photoemission scattering of LaMnO$_3$ far below $T_{\rm OO}$ of 780 K, prior to antiferromagnetic (AFM) transition temperature at the $T_{\rm N}$ of 140 K \cite{Hirota}, were reported \cite{Murakami}. These changes are accompanied by softening of the Raman-active phonon modes \cite{Granado,Laverdiere,Kovaleva_lmo_jpcm}. Similar behaviour arises in another classic example of orbital physics, KCuF$_3$. There the anomalous softening and splitting of the Raman-active modes near $T_{\rm N}$ is related to the existence of nearly degenerate spin and orbital configurations and an unidentified structural transition \cite{Abbamonte}. The evidence of orbital fluctuations accompanied by structure instabilities far below $T_{\rm OO}$, prior to the formation of the magnetic order in these compounds, is not well understood.

Here we used spectroscopic ellipsometry to study in detail the low-energy dielectric function spectra of a detwinned LaMnO$_3$ single crystal in the vicinity of $T_{\rm N}$. The spectra are dominated by the optical band at 2 eV, which is associated with the intersite $e_g-e_g$ electron transition \cite{Kovaleva_lmo_prl,Kovaleva_lmo_prb}. We discovered several narrow satellite excitations above this transition in the 2.2--2.9 eV spectral range. We argue that the discovered excitations could be considered as closely related to the quantum rotor orbital excitations for the $e_g$ electron of Mn$^{3+}$ ions in the potential profile of the ``Mexican hat''  disturbed by the lattice anharmonicity [see Fig. 1(c,d)] \cite{O'Brien}. We had observed some of these excitations (which are obscured by the phonon density-of-states feature in the Raman spectra) in our recent Raman study of LaMnO$_3$ as well, and described them in the cited paper as features of unknown origin \cite{Kovaleva_lmo_jpcm}. Here we also introduce new data on the photoluminescence properties of LaMnO$_3$. The photoluminescence kinetics is characterised by the long discrete lifetimes. We guess that the observed emission occurs from the same energy levels in the 2.2--2.9 eV spectral range, which could be indicative of their electron-vibrational origin. Then, the coupling of $e_g-e_g$ excitation with the quantum rotor states could correspond to the coherent long-lasting exciton propagation, influenced by the anisotropic spin fluctuations around the N\'eel temperature. The existence of orbital excitations is still very controversial and currently is causing many debates in the scientific community due to the lack of conclusive experimental evidence \cite{Kovaleva_lmo_jpcm,Saitoh,Grueninger,Brink}. Here we hope to offer new experimental data in support of this hypothesis. We also suggest that depopulation of the discovered quantum rotor energy levels near the ground state can drive low-temperature structural instabilities, and also facilitate the establishment of the long-range magnetic order in the magnetic compounds with the JT ions.

\section{Experimental approach and results \label{sec:level2}}

\subsection{Crystal growth}
LaMnO$_3$ single crystals were grown by the crucible-free
floating zone method using an image furnace equipped with an
arc lamp \cite{Balbashov,Balbashov1}. The as-grown LaMnO$_3$ single crystals are single phased and exhibit heavily twinned domain patterns in the orthorhombic {\it Pbnm} structure (where the {\bf c} axis is the long axis of the unit cell) at temperatures below the orbital ordering temperature $T_{\rm OO}$\,$\simeq$\,780 K. We were able to remove the twins from an essential part of the sample volume using the procedure described in detail in our previous study \cite{Kovaleva_lmo_prb}. We determined the antiferromagnetic transition temperature at $T_{\rm N}$\,$\simeq$\,139.6 K, which is characteristic of a nearly oxygen-stoichiometric LaMnO$_3$ crystal.

\subsection{Low-energy optical dielectric response of LaMnO$_3$ around the N\'eel temperature}
By using spectroscopic ellipsometry, we investigated the dielectric function spectra, represented by the Kramers--Kronig-consistent real $\varepsilon_{1}(\omega )$ and imaginary $\varepsilon _{2}(\omega )$ parts. The complex dielectric response of LaMnO$_3$ single crystals was investigated in a wide spectral range using a home-built ellipsometer at Max-Planck-Institut f\"ur
Festk\"orperforschung, Germany. The VIS and UV measurements in the
photon energy range of 0.75--6.0\,eV were performed with a home-build ellipsometer of rotating-analyzer type \cite{Kircher}, where the angle of incidence is 70.0$^\circ$. The sample was mounted on a cold finger of a helium flow UHV cryostat where the temperature could be varied between 10 and 300 K. To avoid contamination of the sample surface with ice, we evacuated the cryostat to a base pressure of about $5 \times 10^{-9}$ Torr at room temperature. This method offers significant advantages over conventional reflection methods as (i) it is self-normalizing and does not require reference measurements, and (ii) full complex dielectric response $\varepsilon(\nu)=\varepsilon_1(\nu)+{\rm
i}\,\varepsilon_2(\nu)$ can be obtained directly without a Kramers--Kronig transformation.

Here we report the results of a comprehensive spectroscopic ellipsometry study of a low-energy anisotropic dielectric response of a nearly stoichiometric detwinned LaMnO$_3$ single crystal around the AFM transition temperature $T_{\rm N}$\,$\simeq$\,140 K. Fig. 2(a,b) shows the temperature dependence of the real and imaginary parts of the dielectric function, $\varepsilon(\omega)=\varepsilon_1(\omega)+{\rm i}\varepsilon_2(\omega)$, above the absorption edge in the $\bf a$- and $\bf c$-axis polarisation. The complex dielectric function at low energies is strongly anisotropic and is dominated by a broad optical band at about 2 eV, identifiable by the pronounced resonance and antiresonance features that appear at the same energy in $\varepsilon_2(\omega)$ and $\varepsilon_1(\omega)$, respectively, obeying the Kramers--Kronig relations. Superimposed are a number of smaller spectral features, which look more pronounced in the $\bf a$-axis spectra and less pronounced in the anisotropy-suppressed $\bf c$-axis spectra. 

Figure 2(c,d) shows the corresponding derivative spectra of $\varepsilon_2(\omega)$ over photon energy. These spectra suggest that the low-energy optical band consists of at least three separate optical bands, showing clearly resolved peaks at room temperature at 1.95 $\pm$ 0.01 eV, 2.31 $\pm$ 0.01 eV, and 2.63 $\pm$ 0.01 eV in the $\bf a$-axis polarisation. The bands appear also at slightly different energies in the anisotropy suppressed $\bf c$-axis spectra, where only two peaks at 2.50 $\pm$ 0.01 eV and 2.63 $\pm$ 0.01 eV are visible above the noise level. As follows from Fig. 2(a-d), the anisotropy of the low-energy optical response increases with decreasing temperature, and is accompanied by a pronounced temperature dependence of the constituent optical bands around the N\'eel temperature.

To separate contributions from the higher energy optical bands, which show onset at about 3 eV, and to extract temperature-dependent contribution of the constituent optical bands to the low-energy multi-band structure, we performed a classical dispersion analysis of the temperature-dependent complex dielectric function in the studied spectral range. Using a dielectric function of the form 
$\varepsilon(\omega)=\epsilon_{\infty}+\sum_{j}\frac{S_j\omega_j^2}{\omega_j^2-\omega^2-i\omega\gamma_j}$,
where $\omega_j$, $\gamma_j$, and $S_j$ are  the peak frequency, full width at half maximum (FWHM), and dimensionless oscillator strength of the $j$th oscillator, and $\epsilon_{\infty}$ is the core contribution to the dielectric function \cite{Allen}, we fitted a minimum set of Lorentz oscillators simultaneously to $\varepsilon_1(\omega)$ and $\varepsilon_2(\omega)$. Our fit accurately reproduces the temperature-dependent anisotropic complex dielectric function spectra in the studied temperature range, and shows that the low-energy multi-band structure is represented by the strongly pronounced optical band at about 2 eV and some weaker and narrower higher-energy features, clearly identifiable in the $\bf a$ axis in the room-temperature $\varepsilon(\omega)$ spectra and in the $\bf c$ axis in the low-temperature $\varepsilon(\omega)$ spectra [see Fig. 2(b)].

Figure 3(a-d) shows the results of our dispersion analysis for the low- and room-temperature spectra, where the $\bf a$- and $\bf c$-axis low-energy anisotropic complex dielectric function was represented by the five Lorentz oscillators. Nearly the same resonant energies for the four higher energy features were obtained from the independent fit of the $\bf a$- and $\bf c$-axis $\varepsilon(\omega)$ spectra, so that the peaks appear in the room-temperature $\bf a$-axis spectra at 2.29 $\pm$ 0.02 eV, 2.41 $\pm$ 0.02 eV [these two bands are grouped in a doublet observable at 2.31 $\pm$ 0.01 eV in Fig. 2(d)], 2.66 $\pm$ 0.02 eV, and 2.85 $\pm$ 0.02 eV. In the $\bf c$-axis spectra, we identify the peaks at slightly different energies at 2.25 $ \pm$ 0.02 eV, 2.47 $\pm$ 0.02 eV, 2.66 $\pm$ 0.02, and 2.82 $\pm$ 0.02 eV.

Figure 4(a-c) summarises the $T$-dependence of the peak frequency $\omega_T$, FWHM $\gamma_T$, and oscillator strengths $S_T$ of the strongly pronounced optical band at about 2 eV, resulting from our dispersion analysis of the $\bf a$- and $\bf c$-axis low-energy complex dielectric function. In systems with strong electron--phonon interaction, the shape of an optical absorption band is determined according to the Franck--Condon principle. It is profoundly affected by electron--lattice interaction, which causes both a shift and broadening of the resonance line at elevated temperatures due to lattice anharmonicity effects. When the adiabatic potentials of the ground and excited states are strongly displaced, multiphonon transitions occur with high probability and, thus, a broad intense band in optical absorption is observed. The zero-phonon line in such cases may have a small intensity or may be completely absent. This particular case seems to be relevant to the dominant optical band at about 2 eV, which is strongly pronounced in the $\bf a$-axis polarised spectra and has a FWHM of about 1 eV at low temperature. In the conventional Huang--Rhys theory, the bandwidth has the root-mean-square value $\hbar \omega_0 \sqrt{S}\equiv \hbar \omega_0 \sqrt{S_0{\rm coth}(\hbar \omega_0/k_BT)}$, and the low-temperature limit $\hbar \omega_0\sqrt{S_0}$, where $S_0$ is the Huang--Rhys factor and $\hbar\omega_0$ is the effective phonon frequency. Here the temperature dependence of the FWHM is determined by the temperature dependence of electron--lattice interaction via $S\equiv S_0 {\rm coth} (\hbar \omega_0/k_B T)$ \cite{HayesStoneham}. Taking the low-temperature limit for the FWHM of the 2 eV optical band in the $\bf a$-axis from our data, $\hbar \omega^a_0\sqrt{S^a_0}$\,$\simeq$\,0.96 eV, we were able to accurately fit the FWHM temperature dependence in the 200  -- 300 K temperature range by the formula $\hbar \omega^a_0 \sqrt{S^a_0{\rm coth}(\hbar \omega^a_0/k_BT)}$ and estimate the effective phonon frequency $\hbar \omega^a_0$\,$\simeq$\,27.5 meV [see Fig. 4(b)]. Although the $\bf c$-axis data are much less accurate, especially at the low temperatures due to the  anisotropy signal suppression, we were able to estimate the $\bf c$-axis effective phonon frequency $\hbar \omega^c_0$\,$\simeq$\,29 meV, where the low-temperature limit was $\hbar \omega^c_0\sqrt{S^c_0}$\,$\simeq$\,1.15 eV [see Fig. 4(b)]. The effective phonon frequency $\hbar \omega_0$ determined for the 2 eV charge-transfer excitation can be associated with the longitudinal optical phonon mode in LaMnO$_3$ \cite{Smirnova,Kovaleva_yto_phon_prb}. In the corresponding transverse optical phonon mode, oxygen vibrations in the MnO$_2$ plane modulate total dipole moment along the $\bf a$-axis direction \cite{Smirnova}.

Meanwhile the oscillator strength of the 2 eV optical band exhibits a pronounced increase with decreasing temperature over the entire investigated temperature range in the $\bf a$-axis response, with a noticeable acceleration near $T_{\rm N}$\,$\simeq$\,140 K. The corresponding changes in the $\bf c$-axis are strongly suppressed and exhibit the opposite trend [see Fig. 4(c)]. In our recent optical studies \cite{Kovaleva_lmo_prl,Kovaleva_lmo_prb}, supported by the earlier shell-model calculations 
\cite{Kovaleva_lmo_jetp,Kovaleva_lmo_physB_1,Kovaleva_lmo_physB_2}, we showed an overall consistent picture of the optical spectra and magnetic properties of LaMnO$_3$ in the framework of the superexchange model. These studies  confirm that the low-energy optical transition at about 2 eV is the d--d excitation of the form \mbox{2 Mn$^{3+}$}($t_{2g\uparrow}^3e_{g\uparrow}$) = Mn$^{4+}$($t^3_{2g\uparrow}$) + Mn$^{2+}$($t_{2g\uparrow}^3e^2_{g\uparrow}$), and can be associated with the intersite high-spin Mott-Hubbard excitation of the $e_g$ electron. In agreement with our hypothesis, the $\bf a$- and $\bf c$-components of the oscillator strength, $S_{j}$, and the optical spectral weight (SW) of the 2 eV band shift in an opposite way near $T_{\rm N}$, as influenced by the AFM spin-spin correlations between the Mn$^{3+}$ spins via the superexchange interaction \cite{Kovaleva_lmo_prl, Kovaleva_lmo_prb}. However, as follows from the detailed analysis of the optical SW transfer between the high-spin-(HS) and low-spin-(LS) state transitions in a detwinned LaMnO$_3$ crystal, the $total$ low-energy SW increases in the $\bf a$ axis and decreases in the $\bf c$ axis with decreasing temperature \cite{Kovaleva_lmo_prb}. Below we show that this behaviour could suggest that the ground state is nearly degenerate, and the population of energy levels changes with temperature. 

Suppose that the 2 eV transition actually involves two bands, $A$ and $B$, and that the band $A$ derives from a ground state, which is lower in energy by $\delta$. Suppose also that the band $B$ has a lower intrinsic intensity, e.g. $S_{0B}=fS_{0A}$, where $f<1$. Then the apparent total intensity will be $S_0=\frac{S_{0A}+S_{0B}{\rm exp}(-\delta/k_BT)}{1+{\rm exp}(-\delta/k_BT)}$. From the analysis of the $\bf a$-axis data, we find that $\delta\sim$ 20--30 meV and $f\sim$ 0--0.3 give sensible description of the total intensity $T$-dependence, apart from the region near the $T_{\rm N}$ [see Fig. 4(c)]. As a result, the band $A$ gains its oscillator strength at low temperature, as it arises due to transition from the ground state, whereas the band $B$ looses its oscillator strength, as  the $B$ state is only occupied at rising temperature. In this case,  $\delta\sim$ 20--30 meV will correspond to the estimated effective near-degeneracy of the ground state. We can suggest that the depopulation of the higher energy level with decreasing temperature can result in the orbital structure instability below 250 -- 350 K. This suggestion is in agreement with the onset of the changes observed in the orbital ordering-superlattice reflection near 350 K in X-ray photoemission scattering of LaMnO$_3$ \cite{Murakami}.

At the same time, our dispersion analysis of the $\bf a$-axis dielectric response indicates that the higher energy satellite bands, which appear at room temperature at about 2.29, 2.66, and 2.84 eV, clearly display different behaviour. Contrary to lattice anharmonicity law, their peak positions show a discernible shift, of about 80 meV, to lower photon energies below $T_{\rm N}$. At the same time in the $\bf c$-axis they shift up by a half [see Fig. 4(d,e)]. Additionally, the bandwidths of the two first $\bf a$-axis higher energy satellites, represented by the single Lorentz oscillators, exhibit anomalous temperature dependence. From Fig. 4(f) it can be seen that their bandwidths decrease as temperature increases, showing critical behaviour around $T_{\rm N}$\,$\simeq$\,140 K. This behaviour is also in contradiction to lattice anharmonicity law, which suggests band broadening with rising temperature. The observed behaviour can be understood only if the transitions are represented by two components: (i) one component is wide, corresponding to the low-temperature limit, which disappears at temperatures above approximately 150 K, and (ii) the other component is narrower, corresponding to the high-temperature limit. Meanwhile, in the $\bf c$-axis the bandwidths of these two first higher-energy satellites show an opposite trend, and their FWHMs increase sharply as temperature rises above $T_{\rm N}$\,$\simeq$\,140 K [see Fig. 4(g)]. As a result, the bands appear as sharp features, or may become almost washed out in the $\bf a$-axis and ${\bf c}$-axis spectra, influenced by the anisotropic spin correlations. Similarly to the 2 eV band, the satellite bands gain their oscillator strength in the $\bf a$-axis polarisation, with a discernible kink near $T_{\rm N}$\,$\simeq$\,140 K, whereas the corresponding changes in the $\bf c$-axis reveal an opposite trend [see Fig. 4(h,i)].

And finally, we discuss the behaviour of the 2.8 eV component. The band is relatively wide, of about 0.5 eV, and the bandwidth temperature behaviour in the $\bf a$-axis is seemingly associated with the lattice anharmonicity rule. This association can be described by a large Huang--Rhys factor and some effective phonon frequency. On the other hand, the bandwidth is strongly anisotropic, as it is much smaller in the $\bf c$-axis, where its temperature behaviour is also different [see Fig. 4(f,g)]. In addition, the observed changes in the $\bf a$-axis peak position approaching $T_{\rm N}$ with a discernible down-facing kink below $T_{\rm N}$ [see Fig. 4(d,e)] may be indicative on the complex nature of this band.

Based on the results of our dispersion analysis, we extracted the total temperature-dependent contribution of the four higher-energy satellites, separated from the dominant 2 eV optical transition and from the optical transitions showing up above 3 eV. From Fig. 3(e,f) one can notice a composite structure of the extracted optical response, which exhibits the pronounced temperature dependence near the N\'eel temperature in the $\bf a$ and $\bf c$ axes. Before we discuss the origin of the discovered multiplet excitations, we would like to mention that the observed anomalous temperature behaviour of these excitations around the $T_{\rm N}$ (and in particular, of their bandwidth), influenced by the anisotropic spin correlations [see fig. 4(d-i)], excludes their possible origin associated with any kind of defects, or band-to-band transitions -- of the forbidden or weakly allowed p--d transitions \cite{Moskvin}. Second, the characteristic energy scale of the discovered excitations accompanying the intersite d--d transition at 2 eV and appearing at resonant energies at 2.29 (2.29) eV, 2.41 (2.45) eV, and 2.66 (2.69) eV in  the high- (low-) temperature $\bf a$-($\bf c$-) axis spectra [see Fig. 3(e,f)] cannot be associated with the lattice phonon energies. We considered using additional experimental technique that might help us to explore the nature of these multiplet excitations observed in the ellipsometry spectra in the 2.2--2.9 eV spectral range. Below we present new data we obtained on photoluminescence properties in LaMnO$_3$.

\subsection{Photoluminescence properties of LaMnO$_3$}

The photoluminescence emission spectra of the crystal fixed to a copper holder of a closed-cycle helium refrigerator were measured in right-angle geometry within 10 -- 350 K temperature region using a set-up equipped with a SPM 2 (CARL ZEISS) grating monochromator. Photoluminescence in the 350 --– 870 nm spectral range was detected with a cooled RCA 31034 photomultiplier (GaAs photocathode) operating in the photon-counting mode. Photoluminescence was excited with light from a high-pressure Xe lamp filtered through a double-grating Jobin-Yvon DH 10UV monochromator. All emission spectra taken with spectral resolution of 6 nm were corrected for the spectral dependence of the apparatus response. The excitation spectra are referred to a constant flux of excitation light over the whole spectral region studied.

Time resolved photoluminescence spectrum and photoluminescence decay in the 350--870 nm spectral range was registered using Cary Eclipse fluorescent spectrophotometer. The photoluminescence kinetics dependence was taken with spectral resolution of 20 nm. For the measurements, the samples were mounted on a cold finger\ of a helium-nitrogen flow RTI, Cryomagnetic systems optical cryostat CRYOPT-105.

Figure 5(a) shows the photoluminescence (PL) spectra measured on the same LaMnO$_3$ sample under excitation by light with a photon energy of 4.6 eV at different temperatures from 10 to 300 K. Photoluminescence consists of a broad emission band centered near 2.47 eV at 10 K that shifts to lower energy side with increasing temperature. With the monochromator resolution of 6 nm, the band half-width of about 0.7 eV was observed. The spectra indicate that the broad emission band is actually composed of two nearly unresolved bands as it was really proved in the emission spectrum reconstructed from the PL decay curves [see Fig. 6(b)].

The PL excitation spectra presented in Fig. 5(b) suggest a bulk origin of the emission discovered in LaMnO$_3$ (although there can be also some contribution from surface emission). Indeed, the 4.6 eV excitation corresponds to the resonance energy of the interband O($2p$) -- Mn($3d$) CT transition in a nominally stoichiometric LaMnO$_3$ crystal \cite{Kovaleva_lmo_jetp,Kovaleva_lmo_physB_2}. On the other hand, this excitation is in the range of  the energies of the low-spin states associated with the inter-site $d^4d^4\rightleftharpoons d^5d^3$ transitions between the neighboring Mn$^{3+}$ ions in the lattice. We note that we measured the absorption energies of the  $d^4d^4\rightleftharpoons d^5d^3$ transitions of different symmetry directly by spectroscopic ellipsometry, as shown in our earlier studies [17,18]. According to our data the high-spin (HS) state transition (i) $^6A_1$ appears at 2.0 eV, whereas the low-spin (LS) state transitions (ii) $^4A_1$ and (iii) $^4E_{\varepsilon}$ exhibit close energies around 4.3 eV. In addition, the LS state transition (iv)$^4E_{\theta}$  appears at 4.6 eV, and (v) $^4A_2$ is predicted to appear at higher energies at 6.1 eV.

By assuming that the photoluminescence is a bulk process, the excitation can produce electron-hole pairs, resulting from the CT excitation (a) Mn$^{3+}$+ O$^{2-}$$\rightarrow$ Mn$^{2+}$+ O$^{-}$ and/or (b) 2Mn$^{3+}$ $\rightarrow \ $Mn$^{2+}$+ Mn$^{4+}$. Apparently, the luminescence is due to electron-hole pair recombination. There can be several possibilities for radiative transition. The possibility due to band-to-band recombination, which should involve Stokes shifts of near 1 eV, gives a poor agreement with the position of  the emission band observed at about 2.5 eV. Other possibilities could be due to recombination of electron and hole polarons, as well as radiative emission due to recombination of an exciton. In particular, charge transfer vibronic exciton, involving correlated bound self-trapped electrons and holes, is discussed in relation to the origin of the intrinsic visible band emission peaking at around 2.4 eV (so-called ``green luminescence'') in PbTiO$_3$, SrTiO$_3$, BaTiO$_3$, KNbO$_3$ and KTaO$_3$ perovskite-type crystals \cite{Trepakov}. In our earlier shell model calculations, we used the Mott-–Littleton approach to evaluate polarisation energies in the LaMnO$_3$ lattice associated with holes localised on both the Mn$^{3+}$ cation and the O$^{2-}$ anion. The full (electronic and ionic) lattice relaxation energy for a hole localised at the O site is estimated at 2.4 eV, which is appreciably greater than that of 0.8 eV for a hole localised at the Mn site \cite{Kovaleva_lmo_jetp,Kovaleva_lmo_physB_1}. We think that the PL mechanism due to recombination of the triplet exciton, involving correlated bound self-trapped electrons and a hole, localised on one active O atom and two Mn atoms located on opposite sides from this O atom, is hardly consistent with the large relaxation energy of a hole localised at the O site in LaMnO$_3$. Moreover, due to Coulomb attraction between the electron and hole polarons in the triplet exciton, the recombination energy should be even lower. This analysis indicates that the mechanism associated with recombination of the charge-transfer vibronic exciton \cite{Trepakov} is also seemingly not related to the origin of the photoluminescence in LaMnO$_3$ discovered in the present study.

Exploring the system further experimentally, we were able to detect the time resolved emission spectra for the weak UV-visible photoluminescence in LaMnO$_3$ [see Fig. 6(a-c)]. We investigated the dependence of the PL kinetics under 4.8 eV photon excitation and estimated the lifetimes at some distinct energies of the PL spectrum profile, and the results are presented in Fig. 6(a,b). We found that the PL kinetics is generally characterised by two distinct lifetimes: one is relatively fast, of about 7 ms, whereas the other one is much longer, of about 192 ms. This analysis showed that the taken time resolved PL spectrum is dominated by contribution from the long lifetime process (192 msec).

Surprisingly, we found that the PL emission spectra are very similar to the dielectric function $\varepsilon_2$ spectra of the multiplet transitions in the range 2.2--2.9 eV, extracted in this study from the spectroscopic ellipsometry data [see Figs. 3(e,f) and 5(c)]. However, the discovered PL emission is excited by much higher photon energy of 4.6 eV, and to the higher lying electronic states, seemingly corresponding to the LS states of the intersite $d^4d^4\rightleftharpoons d^5d^3$ transitions. We can suggest that non-radiative relaxation of the electronic excitation occurs from these states to the multiplet energy levels in the range 2.2--2.9 eV, discovered from the present spectroscopic ellipsometry study. These states, appearing in the dielectric function and in the photoluminescence spectra almost at the same energies, could be assigned to the electronic--vibrational (vibronic) excitations of a resonant type addressed below.

\section{Discussion and conclusions \label{sec:level3}}

The double-valued nature of the ``Mexican hat'' potential energy surface [see Fig. 1(c)] means that there are really extra degrees of freedom. Therefore, within any of the JT split doubly degenerate $e_g$ levels of a Mn$^{3+}$ ion, represented  by the ``Mexican hat'' surface, one can have excitations: (i) of conventional phonons, corresponding to the oscillations in the magnitude of the radius $\rho_0=\mid(Q_{20},Q_{30}\mid)$, (ii) of orbitons, or collective oscillations in the ($Q_2$, $Q_3$) plane going along the minimum of the trough, and (iii) where the energy gap between the double-valued energy surfaces matters, 4$\Delta_{\rm JT}$ at the minimum \cite{StonehamTDOS}.

We argue that the multiplet optical excitations discovered above the 2 eV d--d transition [see Fig. 3(e,f)] are closely related to the  orbital excitations of quantum rotor states in the ``Mexican hat'' [see Fig. 1(c)] \cite{O'Brien,StonehamTDOS,Bersuker}. The effect of anharmonicity barriers on the lowest state of the system, which is supposed to derive entirely from the lowest ``Mexican hat'' energy surface, is considered by O'Brien \cite{O'Brien}. The eigenvalues for the {\bf A$_1$}, {\bf A$_2$} and {\bf E} symmetry states are investigated and included in the diagram as a function of the barrier height $\beta$, where also the energy splitting parameter $\alpha$ is used in the quantisation of the discrete energy levels [see Fig.1(d)]\cite{O'Brien}. As follows from the O'Brian's diagram, the $\bf E$  state ($j$=$\pm \frac{1}{2}$) will be the ground state. The eigenvalues for the two extreme cases of the potential barrier, namely for a dynamic fast-rotation limit and a static slow-rotation limit, are given. The orbital excitation energies can be written as $\varepsilon=\alpha j^2$ in the dynamic limit when $\beta=0$. Note that the cubic anharmonicity term only couples states with $j$ values separated by 3. For the ground $\bf E$ state ($j$=$\pm \frac{1}{2}$) the orbital states in the series with $j=\frac{1}{2},-\frac{5}{2},\frac{7}{2},-\frac{11}{2},\frac{13}{2},...$ and $j=-\frac{1}{2},\frac{5}{2},-\frac{7}{2},\frac{11}{2},-\frac{13}{2},...$ are coupled.

In Figure 3(e,f) we plot the energies of the orbital rotor excitations shifted to the high energies with respect to the 2 eV transition [and, in addition, by 70 meV, which is about the energy difference of the levels in the dynamic and static regimes]. The parameter $\alpha$ = 21 cm$^{-1}$ was used, which is in good agreement with the value of the tunnelling splitting $\alpha$ estimated at about 25 cm$^{-1}$ from our recent study of anomalous Raman scattering in LaMnO$_3$ \cite{Kovaleva_lmo_jpcm}. We calculated the following rotor excitation energies corresponding to the modulus of the angular momentum quantum numbers $j$ given in brackets: 0.65 meV (1/2), 16.3 meV (5/2), 31.9 meV (7/2), 78.8 meV (11/2), 110 meV (13/2), 188.1 meV (17/2), 235.0 meV (19/2), 344.3 meV (23/2), 406.8 meV (25/2),  547.4 meV (29/2), 625.5 meV(31/2), ... for the series of the coupled orbital states with $j=\frac{1}{2},-\frac{5}{2},\frac{7}{2},-\frac{11}{2},\frac{13}{2},...$ and $j=-\frac{1}{2},\frac{5}{2},-\frac{7}{2},\frac{11}{2},-\frac{13}{2}$,...\ . Here we note that depopulation of the electron-vibrational levels with energies of about 16 and 32 meV can drive the orbital structure instabilities below about 200 and 380 K, which is in agreement with the observations by Murakami {\it et al.} \cite{Murakami}. We further note that we had observed the excitation at about 80.6 meV (650 cm$^{-1}$) in our recent Raman study of LaMnO$_3$, and described it in the cited paper as a feature of unknown origin \cite{Kovaleva_lmo_jpcm}.

Obviously, the orbital rotor excitation energies are small for the low quantum numbers $j$ (1/2, 5/2, 7/2, 11/2, and 13/2), and cannot be resolved above the 2 eV d--d transition in the present optical experiment. However, for the higher $j$ quantum numbers, the orbital excitations obeying a square law $\varepsilon=\alpha j^2$, can be indeed resolved. We find good correlation of the observed resonant energies of the multiplet structure at  2.29 (2.29) eV, 2.41 (2.45) eV, and 2.66 (2.69) eV in  the high- (low-) temperature $\bf a$-($\bf c$-) axis spectra [see Fig. 3(e,f)], corresponding to the orbital excitations grouped around the higher $j$ values -- (17/2 and 19/2), (23/2 and 25/2), and (29/2 and 31/2) -- which we were able to resolve in our present optical experiment owing to the accuracy of the ellipsometry technique. Unfortunately, the orbital excitations at the higher energies cannot be clearly resolved, as they are superimposed with an intense and relatively wide band at around 2.8 eV, and therefore could be profoundly affected by the behaviour of this band. We would also guess that the orbital rotor excitations probably exist with energies below about $\Delta_{\rm JT}/2$, whereas close to the upper sheet of the ``Mexican hat'' they must disappear. Here we suggest that relatively wide 2.8 eV component could be associated with electronic transition between the double-valued ``Mexican hat'' energy surfaces, of 4$\Delta_{\rm JT}$ at the minimum. Than this will be in agreement with our earlier estimates of JT energy in LaMnO$_3$, $\Delta_{\rm JT}$\,$\simeq$\,0.7 eV \cite{Kovaleva_lmo_prb}.

One can expect that in addition to acoustic orbital wave (Goldstone) mode, resulting from relative motions of different ions, where close ions have similar phases \cite{SarfatStoneham}, there can be also optic orbital modes due to rapid variations in the angle in the ($Q_2$, $Q_3$) plane from one site to the next. Note, if there are potential barriers $\beta$ between the discrete minima on a circle of minimum energy at $\rho_0$, rapid variations in the angle in the ($Q_2$, $Q_3$) plane are only expected to occur with energies near the top of the barriers or higher. Here the lowest states in each minimum are best described by a static JT effect (slow rotation limit), whereas the excited states are best described by a dynamic effect (fast rotation limit). Therefore, we expect that there can be the dynamic orbital wave modes, corresponding to the discovered multiple quantum rotor states in LaMnO$_3$, in the temperature regime above the anharmonicity barriers $\beta$. Here we note that in order of magnitude the barrier's height of the ``Mexican hat'' for Mn$^{3+}$ ions is $\beta\simeq$12 meV (144 K), which is about the N\'eel temperature $T_{\rm N}$\,$\simeq$\,140 K in LaMnO$_3$.

To summarise, in this study of orthorhombic LaMnO$_3$ compound MnO$_6$ complexes were shown to exhibit complex behaviour. On one hand, we see fingerprints of nearly free behaviour of MnO$_6$ complexes within the solid state environment. Our conclusion is based on the observation of the narrow quantised rotational states in LaMnO$_3$, which match the quantum rotor energy level diagram  \cite{O'Brien}. In the dynamic limit when $\beta=0$ these quantum energy levels obey the quadratic dependence $\varepsilon=\alpha j^2$ on the angular momentum $j$ for the series of the coupled orbital states with $j=\frac{1}{2},-\frac{5}{2},\frac{7}{2},-\frac{11}{2},\frac{13}{2},...$ and $j=-\frac{1}{2},\frac{5}{2},-\frac{7}{2},\frac{11}{2},-\frac{13}{2}$,... . The key parameter here is $\alpha$, which is used in the quantisation of the discrete energy levels. The observed narrow quantised rotational states were best described by the parameter $\alpha$ = 21 cm$^{-1}$, which is in good agreement with the value of the tunnelling splitting $\alpha$ estimated at about 25 cm$^{-1}$ from our independent study of anomalous Raman scattering in LaMnO$_3$ \cite{Kovaleva_lmo_jpcm}. On the other hand, we see dynamic-to-static orbital effects, stimulated by the superexchange interaction around the N\'eel temperature. Indeed, we showed that the quantum rotor excitations exhibit the anomalous temperature dependence pronounced near $T_{\rm N}$ and very peculiar for their bandwidth, which disagrees with the lattice anharmonicity law. As a result, they appear as sharp features, or may become almost washed out, corresponding to rotational ordering (static) or nearly free rotational behaviour (dynamic) of the MnO$_6$ complexes, induced by the anisotropic spin correlations. The observed anomalous temperature behaviour of these excitations around the $T_{\rm N}$ (in particular, of their bandwidth), influenced by the anisotropic spin correlations [see fig. 4(d-i)], excludes their possible origin associated with any kind of defects, or band-to-band transitions -- of the forbidden or weakly allowed p--d transitions \cite{Moskvin}. At the same time, the characteristic energy scale of the discovered excitations accompanying the intersite d--d transition at 2 eV and appearing at resonant energies at 2.29 (2.29) eV, 2.41 (2.45) eV, and 2.66 (2.69) eV in  the high- (low-) temperature $\bf a$-($\bf c$-) axis spectra [see Fig. 3(e,f)] cannot be associated with the lattice phonon energies.

Our observation of the quantum rotor states above the d--d excitation at 2 eV, which appear clearly as narrow sidebands, suggests that these states are represented by the coupled d--d exciton and quantum rotors states. These narrow bands appear almost at the same energies in the dielectric function and photoluminescence spectra and could be assigned to the electronic--vibrational (vibronic) excitations of a resonant type.  We guess that the detected long lifetime of photoluminescence kinetics of these coupled exciton--rotor states is indicative of coherent d--d exciton propagation, which may be initiated by transient oscillations at the near resonant frequency of the quantum rotor oscillation. We note that observation of the quantum rotor orbital excitations near the ground state of the lower $e_g$ electronic orbital split by JT effect  in the Raman spectra is difficult, because there they are obscured by the phonon density-of-states feature \cite{Kovaleva_lmo_jpcm}. Observation of the quantum rotor orbital excitations seems to be more straightforward in the optical excitation of the upper $e_g$ electronic orbital. Our present study demonstrates  that spectroscopic ellipsometry and photoluminescence techniques are very useful in studying the properties of the quantum rotor orbital excitations.

Orbital structure instabilities, caused by changes in the
population of the quantum energy levels induced by temperature can result in low-temperature phase transitions, Raman phonon mode anomalies, and facilitate the formation of the long-range magnetic order, or ferroelectric phase in orthorhombic manganites. In order to increase our understanding of these effects in orthorhombic manganites and other systems, where coupling between spin, lattice and orbital degrees of freedom plays an essential role, the nature of the Jahn--Teller ground state and, in particular, the mechanism of the low-temperature phase transitions, associated with temperature populations of the quantum rotor states, need to be studied in more detail. 
 
\ack

We thank A Balbashov for growing the crystals. The authors acknowledge fruitful discussions with D Efremov, G Khaliullin, D Khomskii, B Keimer, M Skorikov, V Vinogradov and N Sibeldin. We also thank A Kulakov for detwinning of the crystal, J Strempfer, I Zegkinoglou and M Schulz for characterisation of the LaMnO$_3$ sample. This work was supported by the Program of the Section of Physical Sciences of the Russian Academy of Sciences ``Strongly correlated electrons in solids and structures'', grant P108/12/1941 of the GACR (Grant Agency of the Czech Republic), grant TA 01010517 of the TACR (Technology Agency of the Czech Republic), and the budget contract N 14.527.11.0004 from 27th, October 2011 of the Ministry of Education and Science of Russian Federation.

\newpage

\begin{figure}
\includegraphics*[width=140mm]{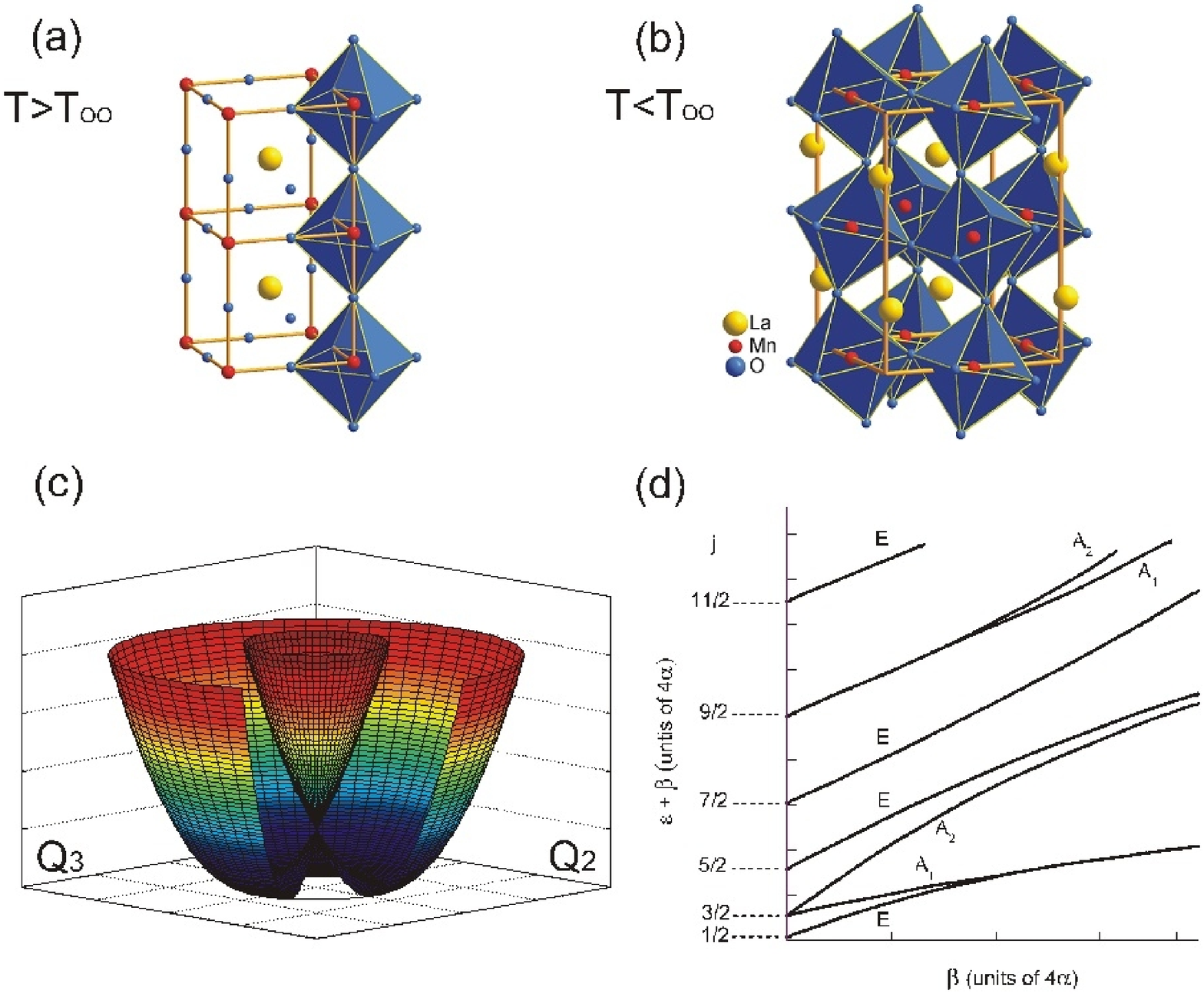}
\caption{(a) Structure of high-temperature perovskite phase. (b) Structure of orthorhombic phase below orbital ordering temperature $T_{\rm OO}$. (c) Double-valued ``Mexican hat'' potential energy surface in a clean dynamic limit. The surface is a parabola of revolution. The ``Mexican hat'' deformed by anharmonicity has three discrete minima at the bottom of the trough, which are separated by anharmonicity barriers (not shown). (d) The energies in the ``Mexican hat'' as a function of the barrier parameter $\beta$. Values of the angular momentum $j=\frac{1}{2}$,$\frac{3}{2}$,$\frac{5}{2}$,... are given in a dynamic limit. Results after O'Brien \cite{O'Brien}.}
\end{figure}

\newpage

\begin{figure}
\includegraphics*[width=80mm]{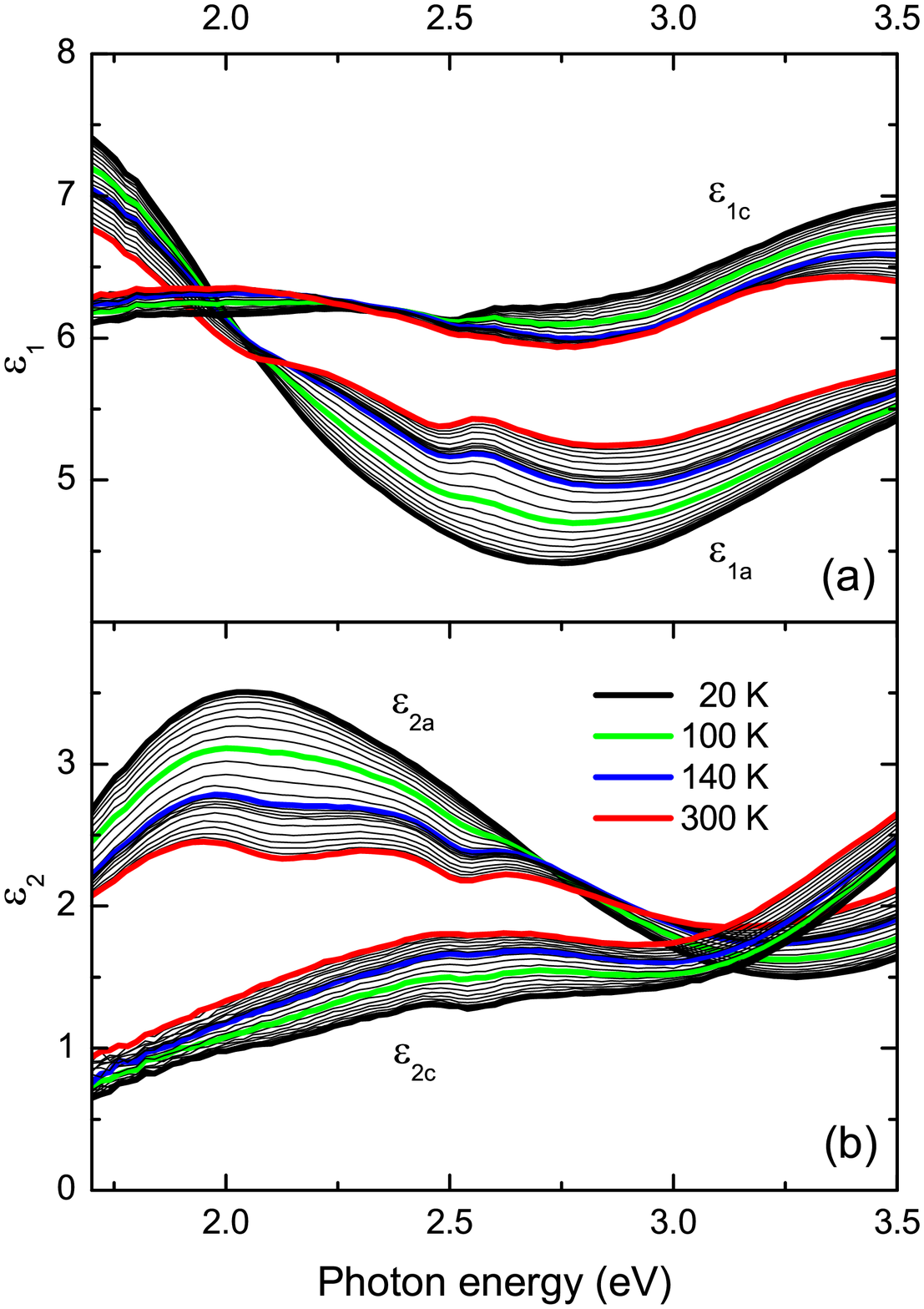}
\includegraphics*[width=80mm]{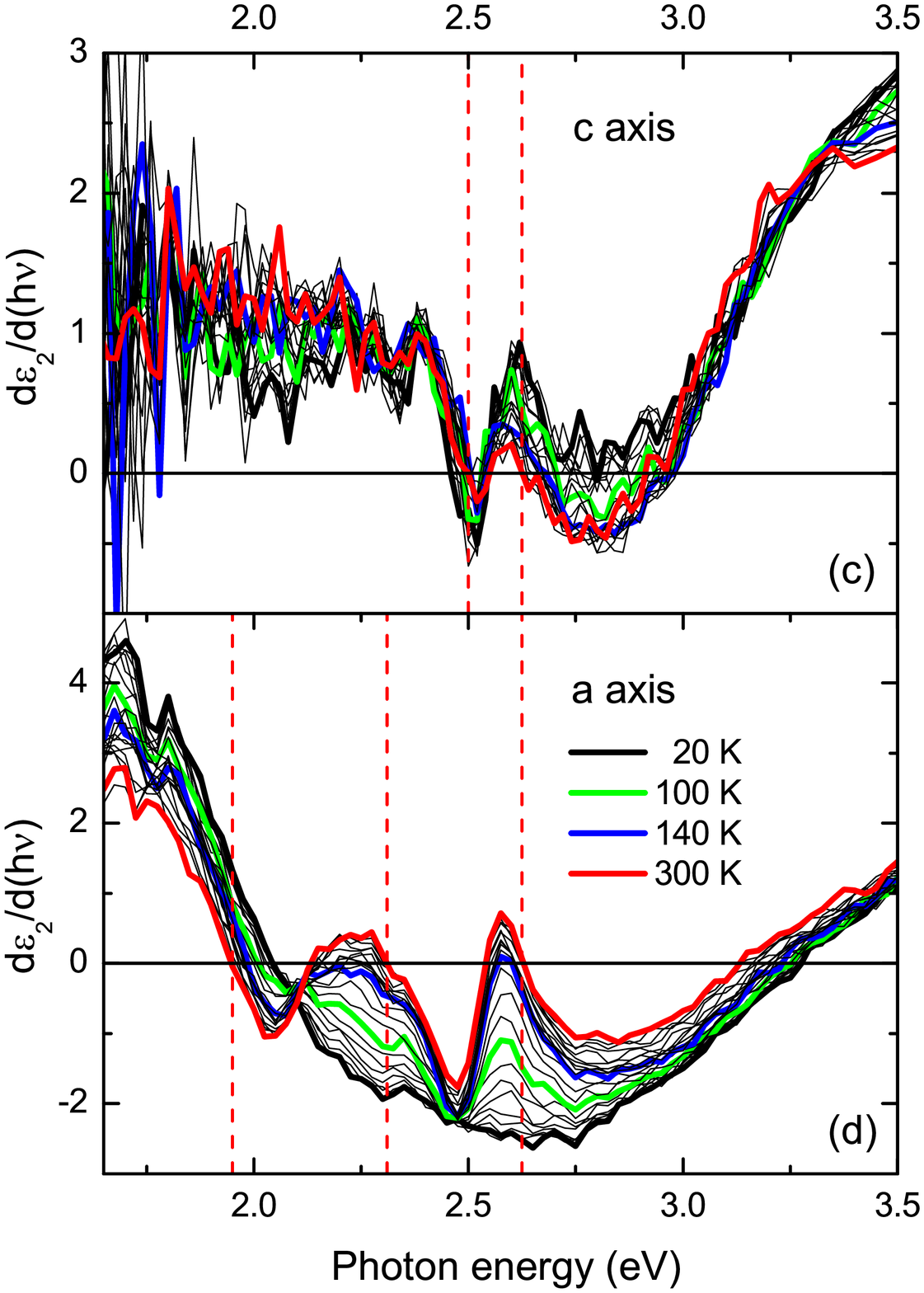}
\caption{Temperature variation of (a) real $\varepsilon_1(\omega)$ and (b) imaginary $\varepsilon_2(\omega)$ parts of the complex dielectric function spectra of the detwinned LaMnO$_3$ crystal in the $\bf a$-axis and $\bf c$-axis polarisation (of the $Pbnm$ symmetry unit cell). The representative spectra at the temperatures around $T_{\rm N}\simeq140$ K are indicated. The temperature evolution is shown in successive temperature intervals of 10 K between 20 and 160 K and of 25 K between 175 and 300 K. (c,d) The derivative of the measured $\varepsilon_2(\omega)$ spectra over photon energy in the $\bf a$ axis and $\bf c$ axis. The energies of zero crossing, corresponding to the peak positions in the $\varepsilon_2(\omega)$ spectra at $T$ = 300 K, are marked by the red dashed lines.}
\end{figure}

\newpage

\begin{figure}
\includegraphics*[width=51mm]{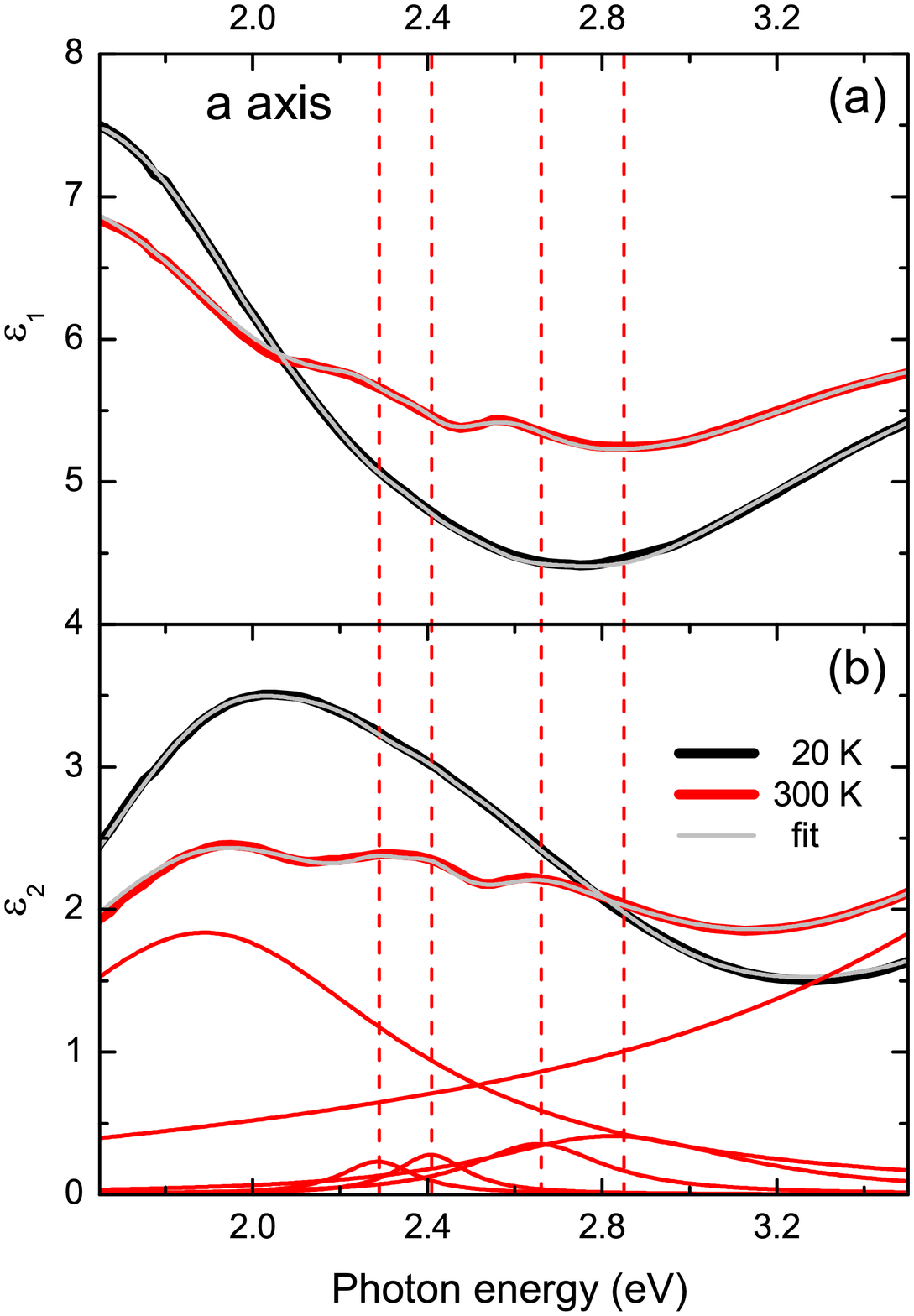}
\includegraphics*[width=51mm]{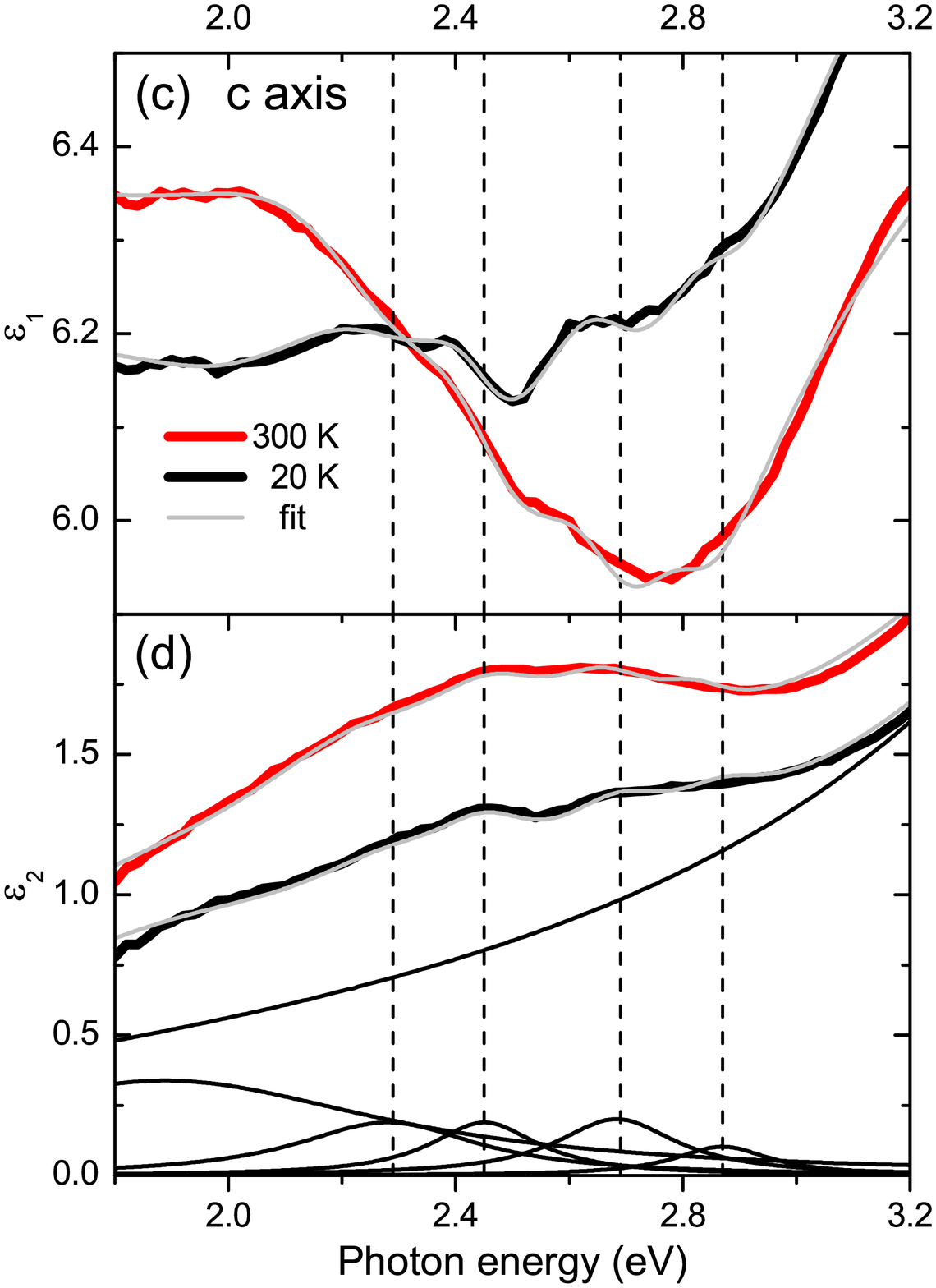}
\includegraphics*[width=51mm]{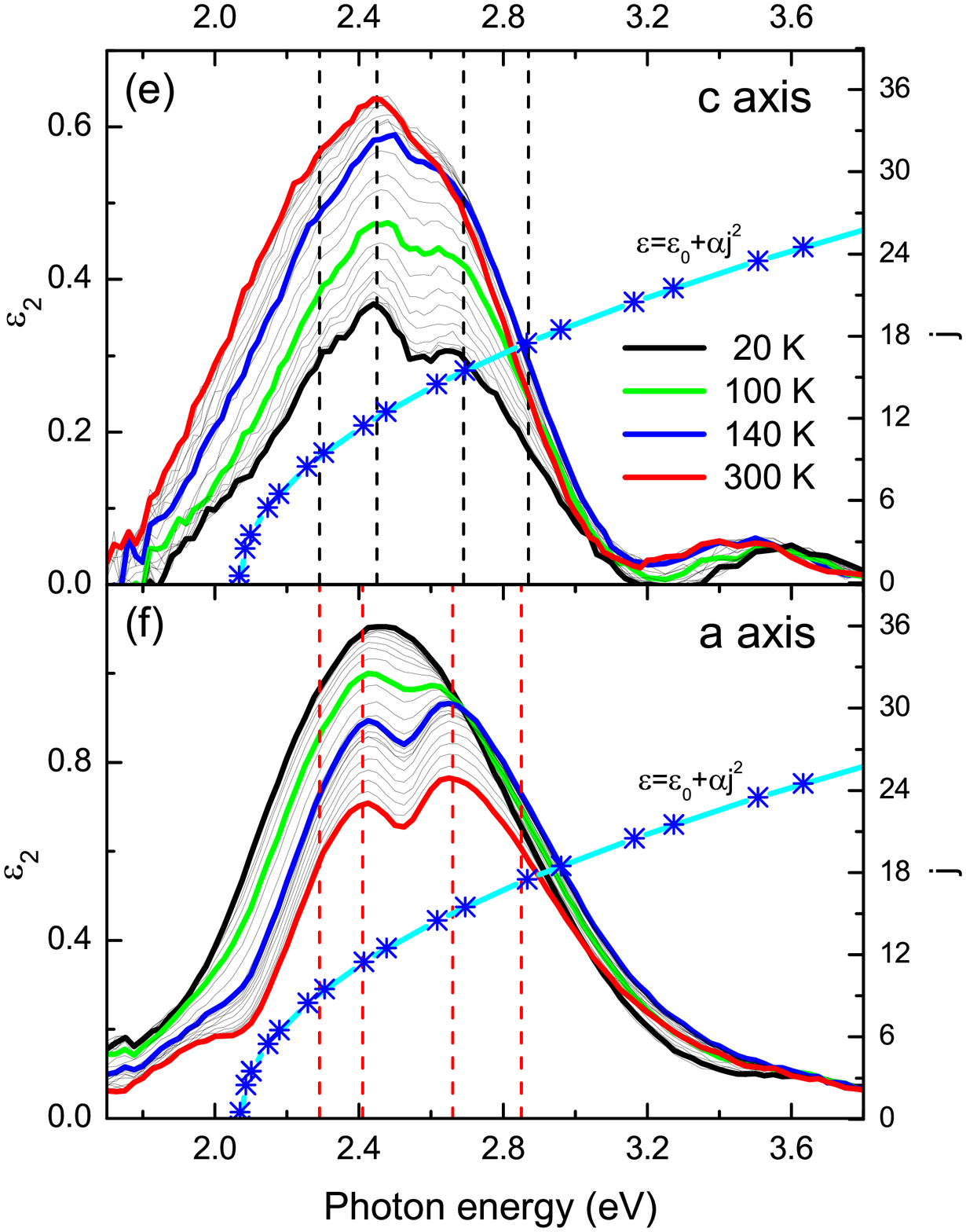}
\caption{(a,c) Real $\varepsilon_1(\omega)$ and (b,d) imaginary $\varepsilon_2(\omega)$ parts of the $\bf a$- and $\bf c$-axis dielectric function spectra $\varepsilon(\omega)$ at 20 K and at 300 K, fitted by the Lorentzian oscillators according to the classical dispersion analysis, as described in the text. The five Lorentzian bands determined from the fit of the $\varepsilon(\omega)$ spectra in the {\bf a} axis at 300 K and in the {\bf c} axis at 20 K are drawn; the corresponding resonant energies of the four higher-energy satellites are marked by the dashed lines. (e,f) The temperature dependence of the four higher energy excitations in the {\bf a}-axis and {\bf c}-axis spectra, resulting from the dispersion analysis. Calculated energies of the quantum rotor levels, $\varepsilon=\alpha j^2$ where $\alpha=21$ cm$^{-1}$, are shown with the star symbols, shifted up by the energy of the d--d transition at 2 eV and by the energy difference  of the levels in the dynamic and static regimes of about 70 meV.}
\end{figure}

\newpage

\begin{figure}
\includegraphics*[width=100mm]{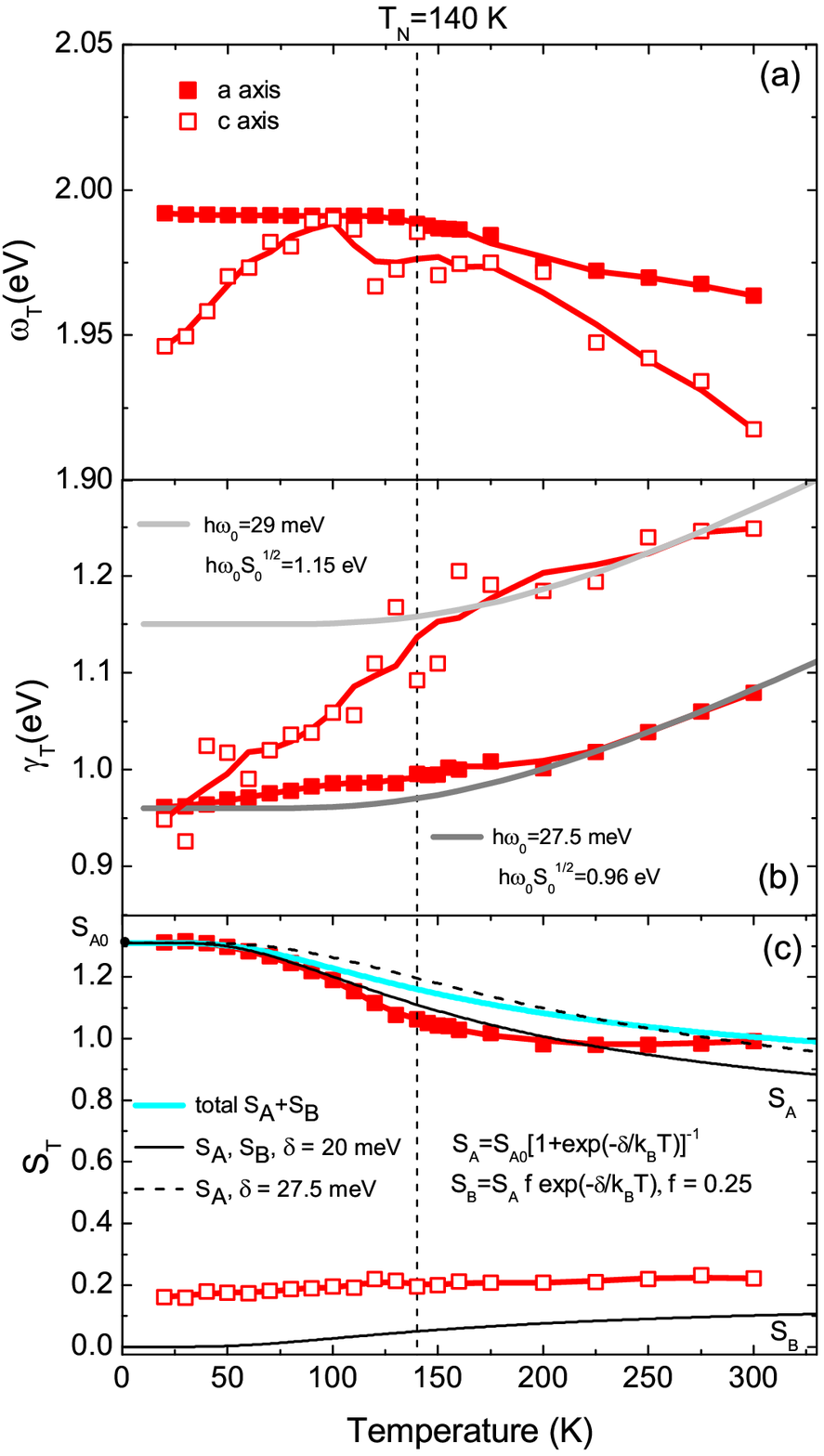}
\hspace{-9.2em}\includegraphics*[width=100mm]{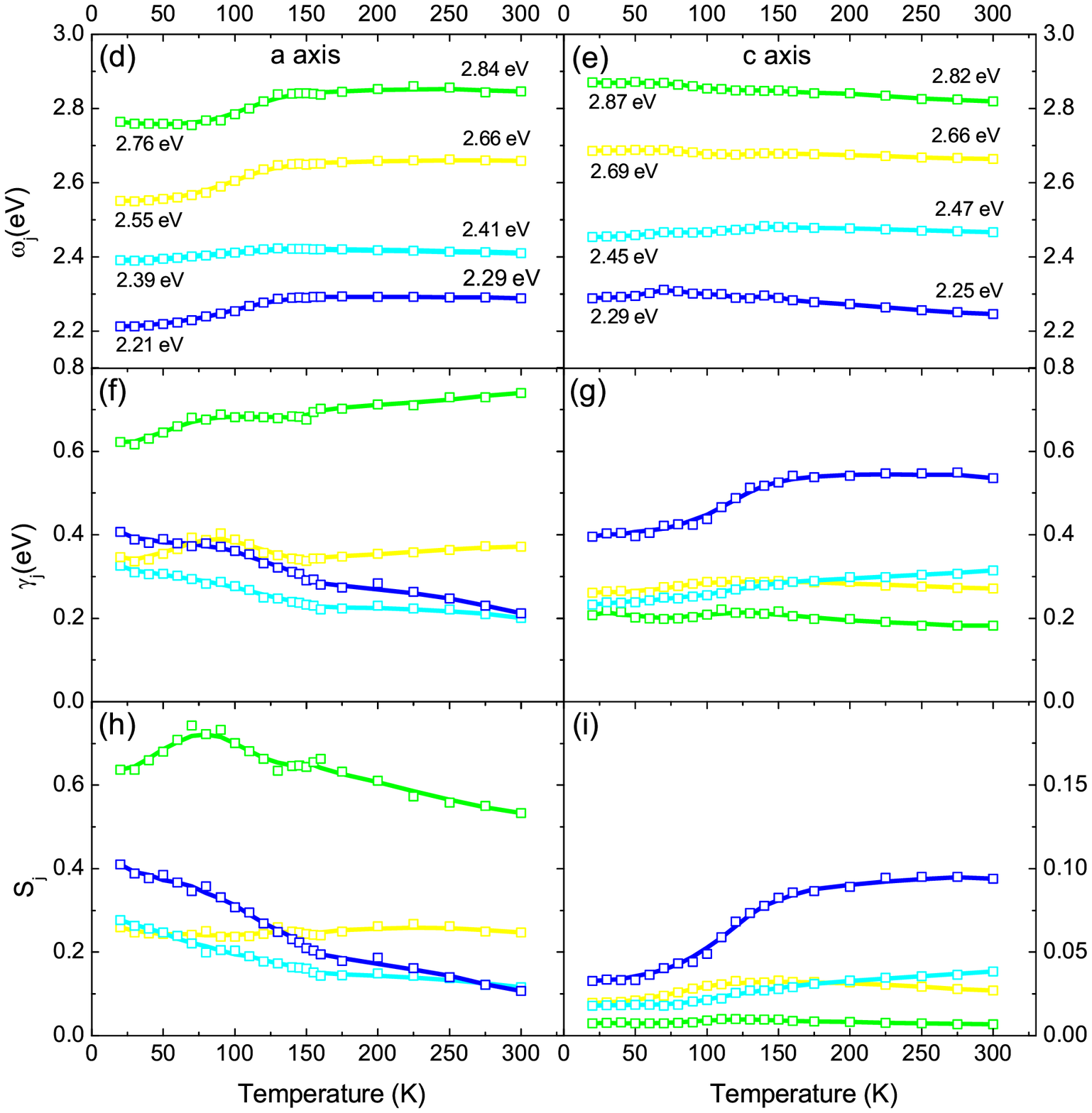}
\vspace{-2.5cm}\caption{Temperature dependence of the peak energies $\omega_j$, full widths at half maximum (FWHMs) $\gamma_j$, and oscillator strengths $S_j$ for the dominating 2 eV band (a,b,c) and for the four higher-energy satellites in the $\bf a$-axis (d,f,h) and in the $\bf c$-axis (e,g,i) polarisation, resulting from the dispersion analysis, as described in the text.}
\end{figure}

\newpage

\begin{figure}
\hspace{-0.7cm}\includegraphics*[width=56mm]{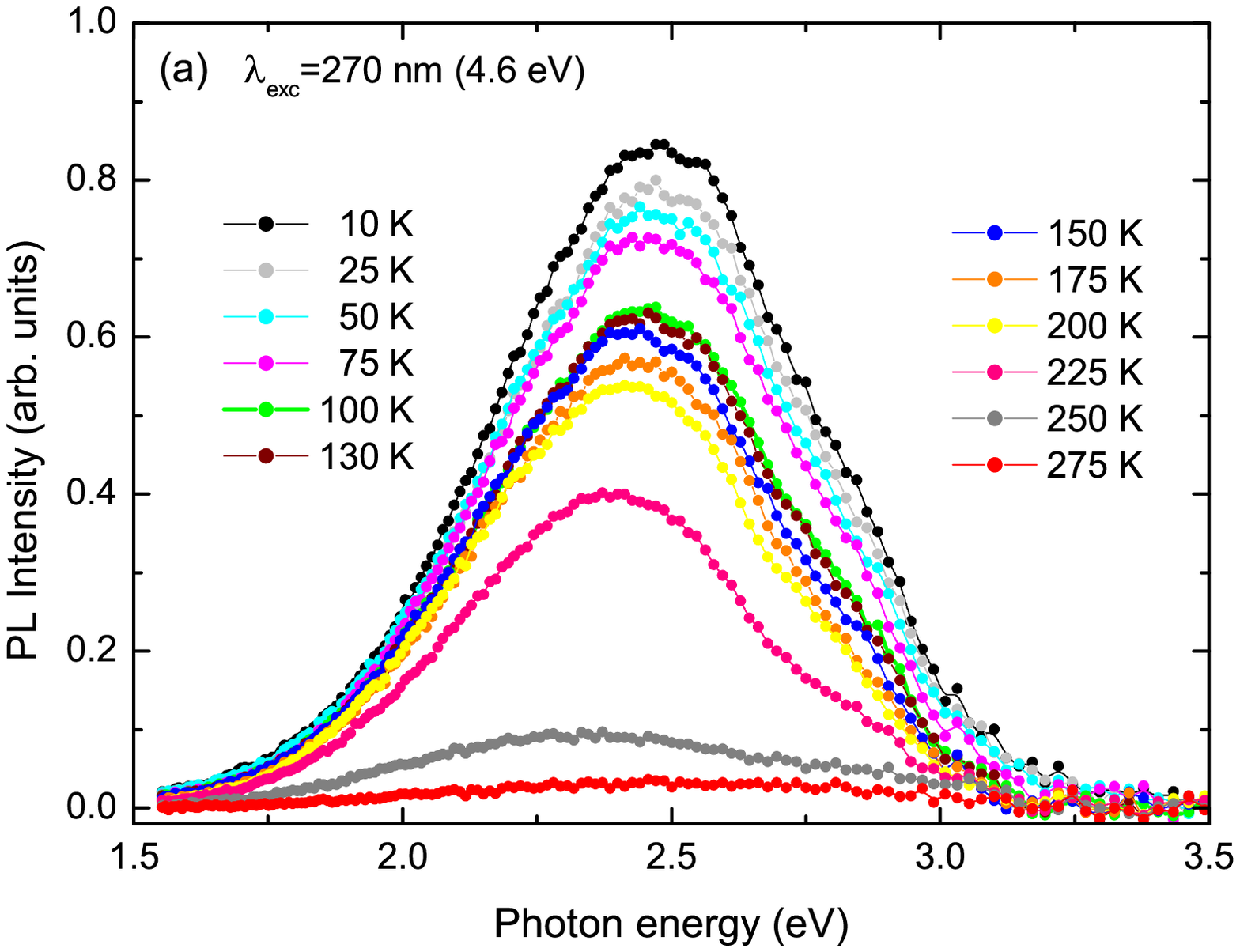}
\hspace{-0.4cm}\includegraphics*[width=56mm]{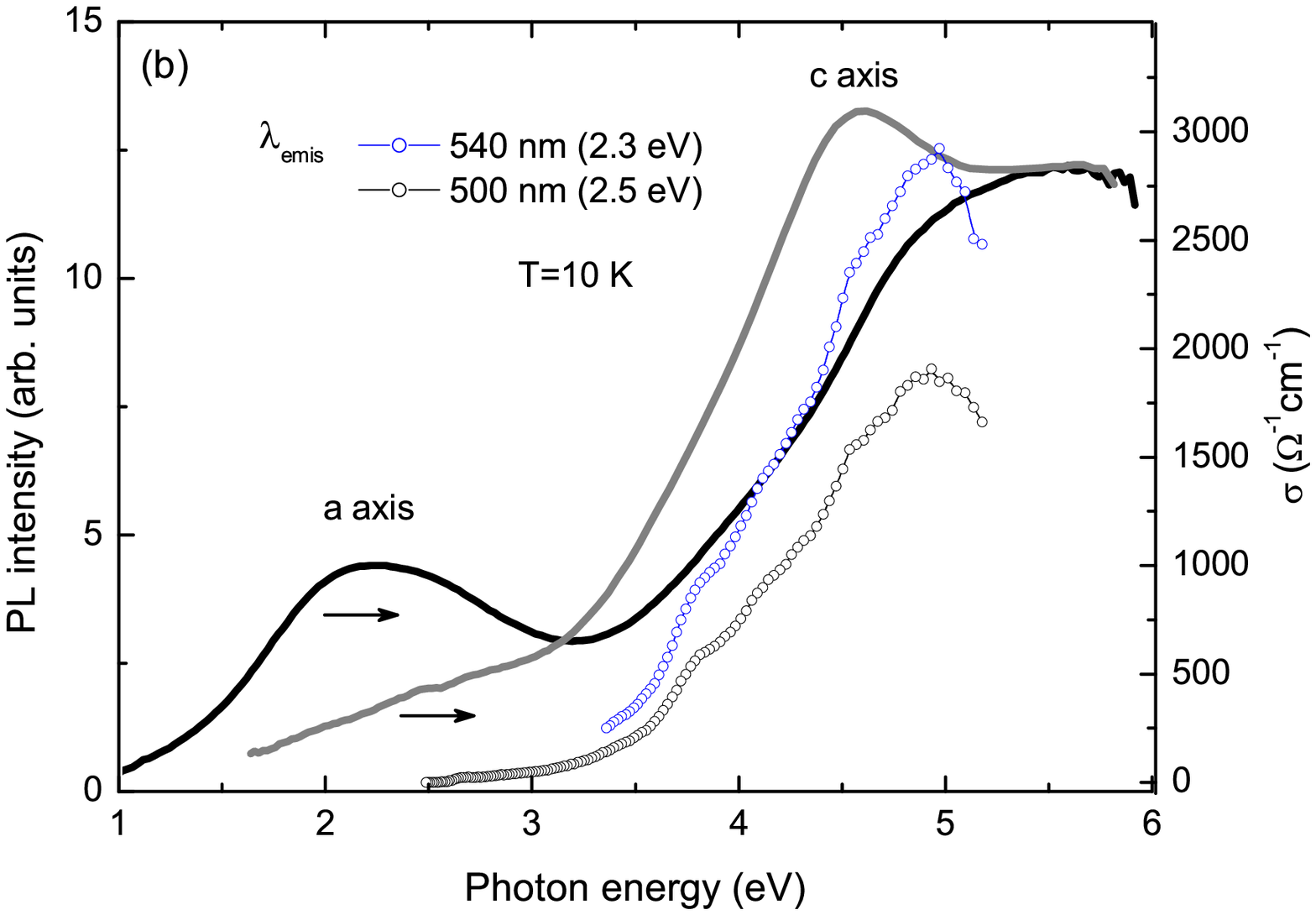}
\hspace{-0.4cm}\includegraphics*[width=56mm]{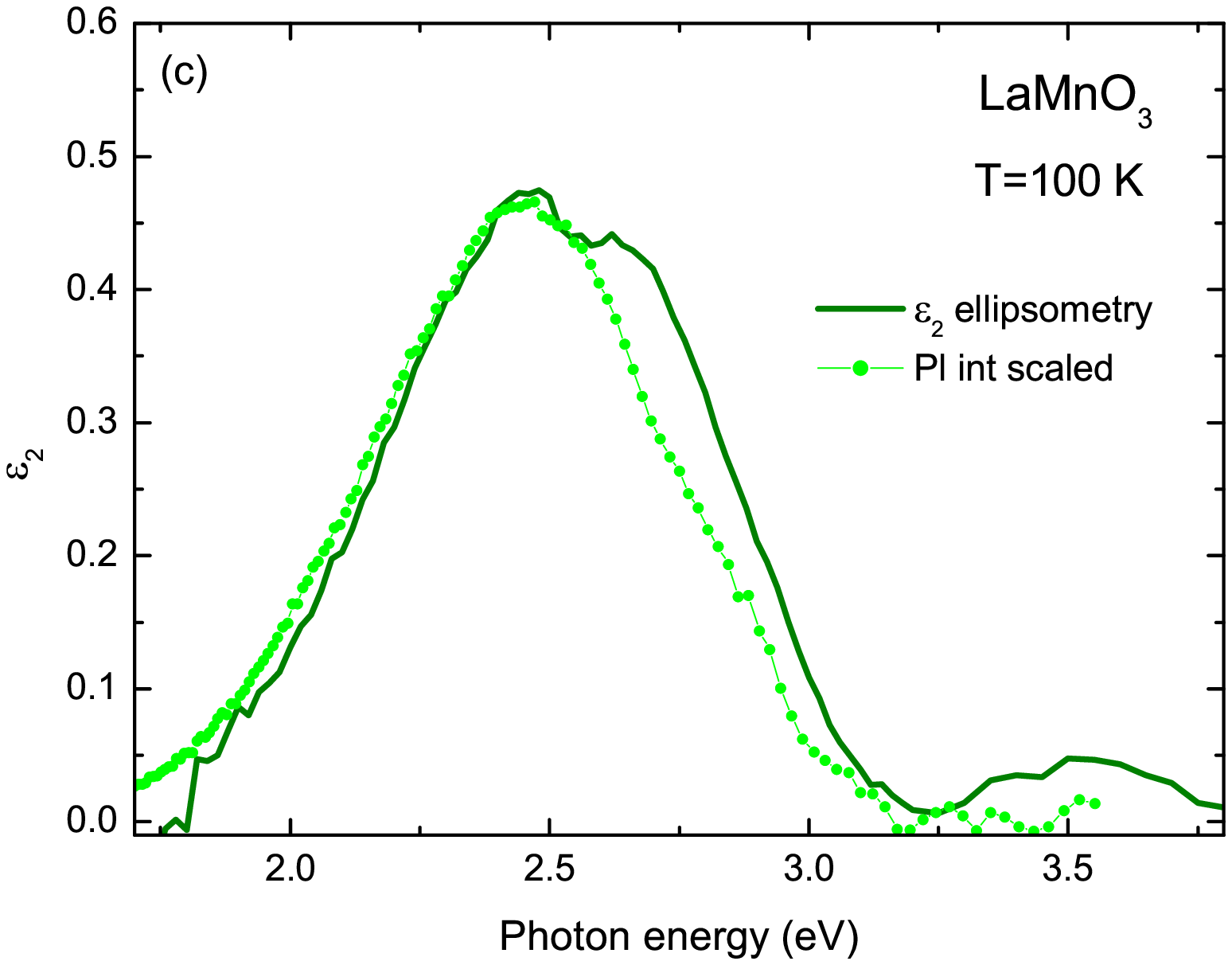}\\
\vspace{-1cm}\caption{(a) Photoluminescence emission spectra in a broad temperature range across the phase transition temperature $T_{\rm N}$\,$\simeq$\,140 K. (b) Photoluminescence  excitation spectra, superimposed with the wide-range anisotropic optical conductivity spectra $\sigma^{\rm {\bf a,c}}(\omega)$. (c) The extracted contribution of the four higher-energy excitations to the $\varepsilon_2$ dielectric function in the {\bf c}- axis, resulting from the dispersion analysis, compared with  the scaled intensity of the photoluminescence spectrum, measured at 100 K.}
\end{figure}

\newpage

\begin{figure}
\hspace{-1.2em}\includegraphics*[width=170mm]{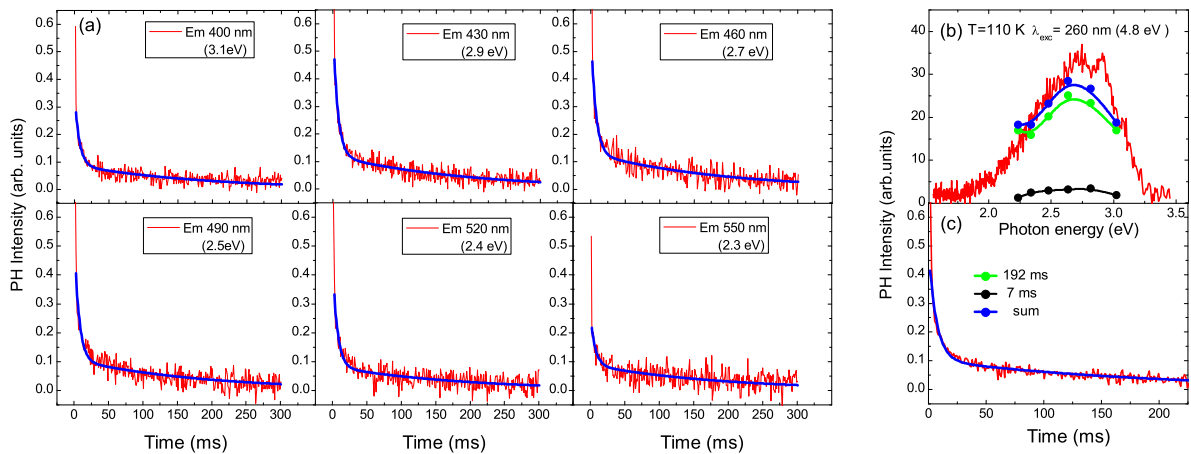}
\vspace{-16cm}\caption{(a) Kinetics of photoluminescence under excitation by photon energy of 4.8 eV, measured at the descrete energies of the emission spectrum. (b) The photoluminescence spectrum of LaMnO$_3$, represented by relative contribution of the emission with two characteristic lifetimes. (c) Summary response of the kinetics presented in (a), used in the fitting of the lifetimes.}
\end{figure}

\end{document}